\documentclass[a4paper,12pt]{article}
\usepackage[T1]{fontenc}
\usepackage{graphicx}
\usepackage{bm}
\usepackage{authblk}
\usepackage{xcolor}
\usepackage{subcaption}
\usepackage{amsmath}
\usepackage{amsfonts}
\usepackage{setspace}
\usepackage[margin=1in]{geometry}
\usepackage{tikz}
\usepackage{cite}
\usepackage{url}
\usepackage[font=small]{caption}
\newcommand\up[1]{\textsuperscript{#1}}

\title{Machine learning analysis of rogue solitons in supercontinuum generation}

\author[1,*]{\small Lauri Salmela}
\author[2]{\small Coraline Lapre}
\author[2]{\small John M. Dudley}
\author[1]{\small Go{\"e}ry Genty}

\affil[1]{\footnotesize Photonics Laboratory, Tampere University, Tampere, FI-33014, Finland}
\affil[2]{\footnotesize Institut FEMTO-ST, Universit\'{e} Bourgogne Franche-Comt\'{e} CNRS UMR 6174, Besan\c{c}on, 25000, France}
\affil[*]{lauri.salmela@tuni.fi}

\date{}

\begin{document}

\maketitle

\begin{abstract}
Supercontinuum generation is a highly nonlinear process that exhibits unstable and chaotic characteristics when developing from long pump pulses injected into the anomalous dispersion regime of an optical fiber. A particular feature associated with this regime is the long-tailed ``rogue wave''-like statistics of the spectral intensity on the long wavelength edge of the supercontinuum, linked to the generation of a small number of ``rogue solitons'' with extreme red-shifts. Here, we apply machine learning to analyze the characteristics of these solitons at the edge of the supercontinuum spectrum, and show how supervised learning can train a neural network to predict the peak power, duration, and temporal delay of these solitons from only the supercontinuum spectral intensity without phase information. The network accurately predicts soliton characteristics for a wide range of scenarios, from the onset of spectral broadening dominated by pure modulation instability to near octave-spanning supercontinuum with distinct rogue solitons.   

\end{abstract}

\section*{Introduction}

The broadband fiber supercontinuum (SC) has developed into a widespread and versatile light source that has found many important applications in areas such as spectroscopy, imaging, and precision frequency metrology \cite{dudley2006}. The noise properties of broadband SC spectra have been a subject of much interest, not only from an applications perspective, but also as an example of the rich instability dynamics that can arise in a highly nonlinear system.  Of particular focus has been been the study of SC generation seeded by picosecond (or longer) pulses, where the spectral broadening is triggered by modulation instability (MI) which exponentially amplifies noise components outside the pump spectral bandwidth \cite{dudley2006}. In the time domain, this instability induces the break-up of the pump pulse envelope into a series of high-intensity breathers with random characteristics \cite{dudley2009}.  With further propagation and the influence of higher-order dispersion and stimulated Raman scattering, these temporal breathers evolve into localized soliton structures whose subsequent dynamics seed the development of a broadband SC spectrum \cite{dudley2006,genty2007}. The noise-seeded nature of the overall process leads to an incoherent spectrum with large shot-to-shot spectral fluctuations particularly pronounced on the long wavelength edge \cite{dudley2006,solli2007}.

There has been much interest in studying the spectral fluctuations on the long-wavelength edge of the broadband SC \cite{solli2007}, as these have been shown to exhibit highly skewed statistics associated with extreme events and the emergence of rogue solitons of tens of femtosecond duration \cite{solli2007,Mussot2009,erkintalo2009,erkintalo2010,dudley2014}.  Originally characterized using long-pass filtering and the real-time dispersive Fourier transform \cite{solli2007}, the observation of these fluctuations stimulated a large number of subsequent theoretical, numerical and other experimental investigations, linking the emergence of rogue solitons with collision dynamics during the initial SC development phase \cite{Mussot2009,erkintalo2010,Antikainen2012}.  Related work has since extended the study of the properties of rogue solitons to a wider range of SC generation regimes and input conditions \cite{dudley2008, erkintalo2009, wetzel2012, godin2013, akhmediev2016}.  

The dispersive Fourier transform technique has now become a standard tool to study  ultrafast instabilities in the spectral domain \cite{Goda2013,mahjoubfar2017}, yielding significant insight into the dynamics of many complex nonlinear systems \cite{solli2012, wetzel2012, godin2013, herink2017, krupa2017, ryczkowski2018,lapre2019}. However, a drawback of the dispersive Fourier transform  is that it does not provide information on the spectral phase, and thus does not allow a quantitative analysis of the associated temporal intensity profiles.  In order to characterize real-time temporal fluctuations associated with incoherent dynamics, more complex techniques using time-lens or heterodyne approaches need to be used \cite{narhi2016,suret2016,ryczkowski2018,tikan2018,liu2018}, yet measurements in this case are generally restricted to specific (narrow bandwidth) wavelength ranges with sub-ps timescale resolution.  These limitations preclude the characterization of solitons or localized structures with 10's of femtosecond duration.

In this paper, we show how machine learning (ML) can address this problem by analyzing real time spectral intensity measurements in a way that allows key temporal characteristics of SC rogue solitons to be determined.  More specifically, we train a supervised neural network (NN) using numerical simulations of the generalized nonlinear Schr{\"o}dinger equation (GNLSE) to correlate key temporal characteristics  (peak power, duration, temporal delay) of the most red-shifted rogue soliton in a SC with the corresponding complex supercontinuum spectral intensity profile.  Despite the absence of any phase information at the network input, the trained network is able to infer the red-shifted soliton characteristics with excellent accuracy, exceeding that obtained when using a single-parameter metric such as e.g. the spectral bandwidth or energy in the long-wavelength edge of the SC spectrum. Significantly, the NN analysis remains accurate over a wide variety of dynamical regimes,  from the onset of spectral broadening dominated by modulation instability, to near octave-spanning supercontinuum with distinct rogue solitons.   


\section*{Results}

\subsection*{Rogue soliton generation and machine learning}

\begin{figure}[t]
  \begin{center}
  \includegraphics[width=\textwidth]{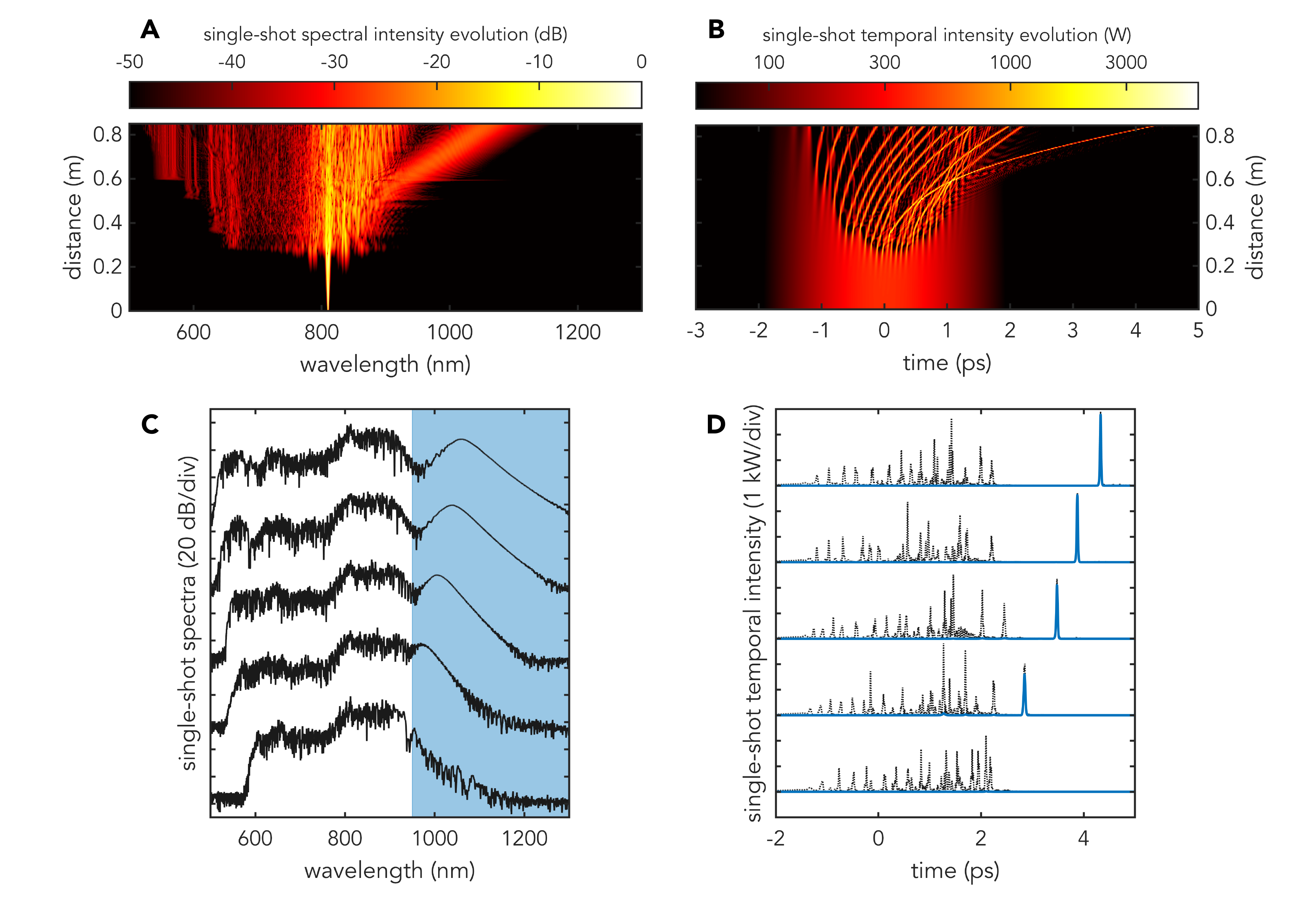}
  \end{center}
  \caption{Dynamics of noisy supercontinuum generation. Single-shot temporal (A) and spectral (B) evolution where we see enhanced spectral broadening associated with the emergence of a rogue soliton. (C) and (D) show a selection of 5 random supercontinuum spectra and corresponding temporal intensity profiles in black. The blue curve in (D) shows the rogue soliton temporal intensity associated with the long-pass filtered spectral components (blue area) in (C) while the black dotted line shows the full time-domain intensity.}
  \label{fig:SCdyn}
\end{figure}

We begin by illustrating in Fig.~\ref{fig:SCdyn} the noise sensitivity of SC generation and rogue soliton emergence when spectral broadening is seeded from a long pump pulse injected into the anomalous dispersion regime of a highly nonlinear fiber. The input conditions here correspond to hyperbolic-secant pulses with 2 ps duration (full-width at-half-maximum FWHM), 400~W peak power and 810~nm central wavelength injected into an 85~cm long photonic crystal fiber (PCF) with zero-dispersion at 785~nm. Input noise is included in the spectral domain using a standard one-photon-per-mode background with random phase \cite{dudley2006}. See Methods for additional simulation parameters.  

We performed simulations to generate an ensemble of 30,000 pairs of spectra and temporal intensity profiles using different random noise seeds.  The top panels in Fig.~\ref{fig:SCdyn} plot results from one realization in the ensemble to show the spectrum (A) and the corresponding temporal intensity profile (B) associated with rogue soliton characteristics.  In particular, we clearly identify the initial stage of noise-seeded modulation instability after a propagation distance of about 30~cm, with the generation of distinct spectral sidebands and the break-up of the pump pulse envelope into multiple breather structures \cite{dudley2009}. With further propagation, the breathers evolve into fundamental solitons experiencing Raman self-frequency shift and  separating in time from the residual background envelope \cite{Mahnke2012}. After a distance of $\sim$50 cm, collisions between multiple solitons increase the frequency-shit rate of one of the solitons emerging from the envelope center, leading to the generation of a distinct rogue soliton with extreme red-shift \cite{Mussot2009,erkintalo2010,Antikainen2012}. 

The results in (C) show a  selection of 5 supercontinuum spectra from the ensemble, and the black lines in (D) show the corresponding temporal intensity profiles.  Note that we have sorted the results in (C) and (D) to show increasing spectral width from top to bottom but of course these results occur randomly in the ensemble.  From these results, we clearly see how the noise-sensitive dynamics lead to large shot-to-shot spectral variations between different results in the ensemble, and we see particularly how the shot-to-shot spectral variations in the long-wavelength edge correspond to isolated solitons with different peak power, duration and temporal delay. 

The approach previously used in experiments \cite{solli2007} to isolate these rogue solitons was to apply the dispersive Fourier transform for real-time spectral characterization in conjunction with a long-pass spectral filter that transmits only spectral components beyond the filter cut-on wavelength. The numerical results in Fig.~\ref{fig:SCdyn}C-D shows how this filtering indeed isolates the temporal solitons, where the blue region in C shows the filter cutoff region, and the blue curves in D shows the corresponding temporal intensity profiles after numerical Fourier transform.  However, the temporal duration of these rogue solitons (10's of femtoseconds) is too short for direct real-time measurements.

\begin{figure}[t]
  \begin{center}
  \includegraphics[width=\textwidth]{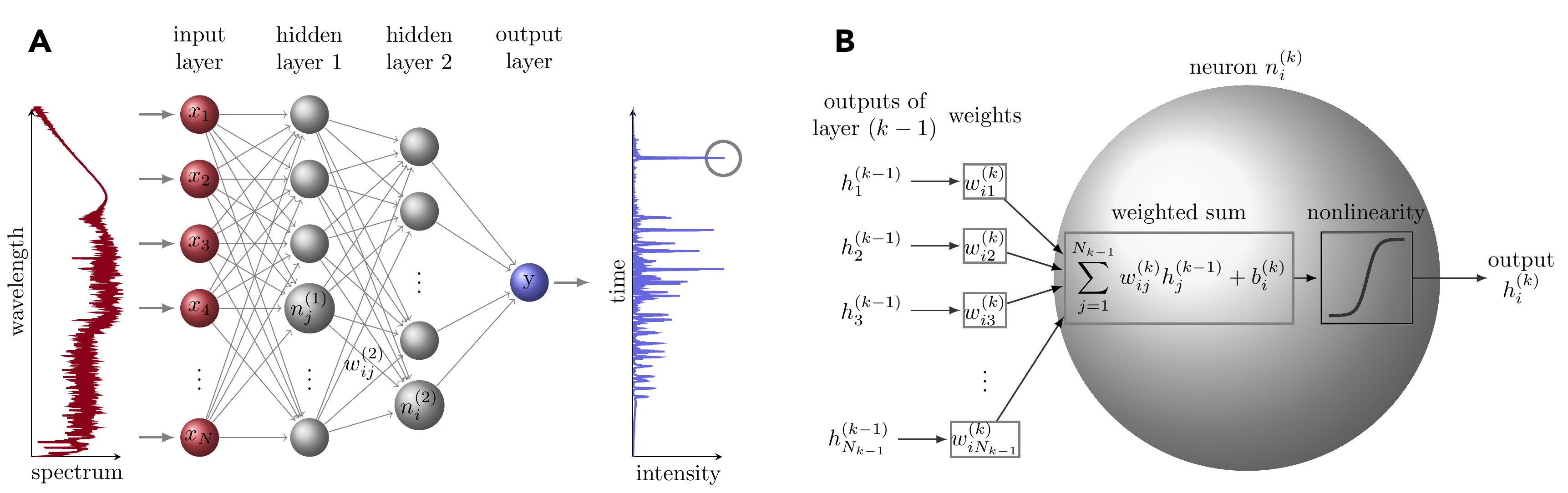}
  \end{center}
  \caption{Schematic of the feed-forward neural network used in this work. (A) The input of the network is a single-shot supercontinuum spectral intensity vector ${\bf X}_n = [x_1, x_2, \cdots , x_N]$ yielding the output of the network $y$ that corresponds to the peak power or temporal shift of the rogue solitons. The network consists of two fully connected hidden layers and a single output neuron. (B) shows the operation of a single neuron. The output of a generic neuron $n_i^{(k)}$ ($i$th neuron in layer $k$) is calculated as a weighted sum between the outputs from the previous layer $k-1$ and the weights of each connection $w_{ij}^{(k)}$ which is followed by adding a bias term $b_i^{(k)}$ and nonlinearity. See Methods for more details.}
  \label{fig:NN}
\end{figure}

But it is here that we can apply machine learning to extract the temporal characteristics of the rogue solitons only from simple unfiltered spectral measurements.  Specifically, we apply a supervised feed-forward neural network to correlate the full spectrum and the temporal characteristics of the most red-shifted soliton of a given (noisy) SC spectrum. A schematic of the neural network is shown in Fig.~\ref{fig:NN}. The NN input is a vector (${\bf X}_n = [x_1, x_2, \dots , x_N]$) corresponding to the SC spectrum, whilst the NN output is a scalar value equal to the maximum temporal intensity, duration or temporal delay (defined with respect to the pump pulse center) of the most red-shifted soliton. The NN is trained from an ensemble of 20,000 simulated single-shot SC using a conjugate gradient back-propagation method \cite{fletcher1964}. See Methods for further details. 

\begin{figure}[t]
  \begin{center}
  \includegraphics[width=\textwidth]{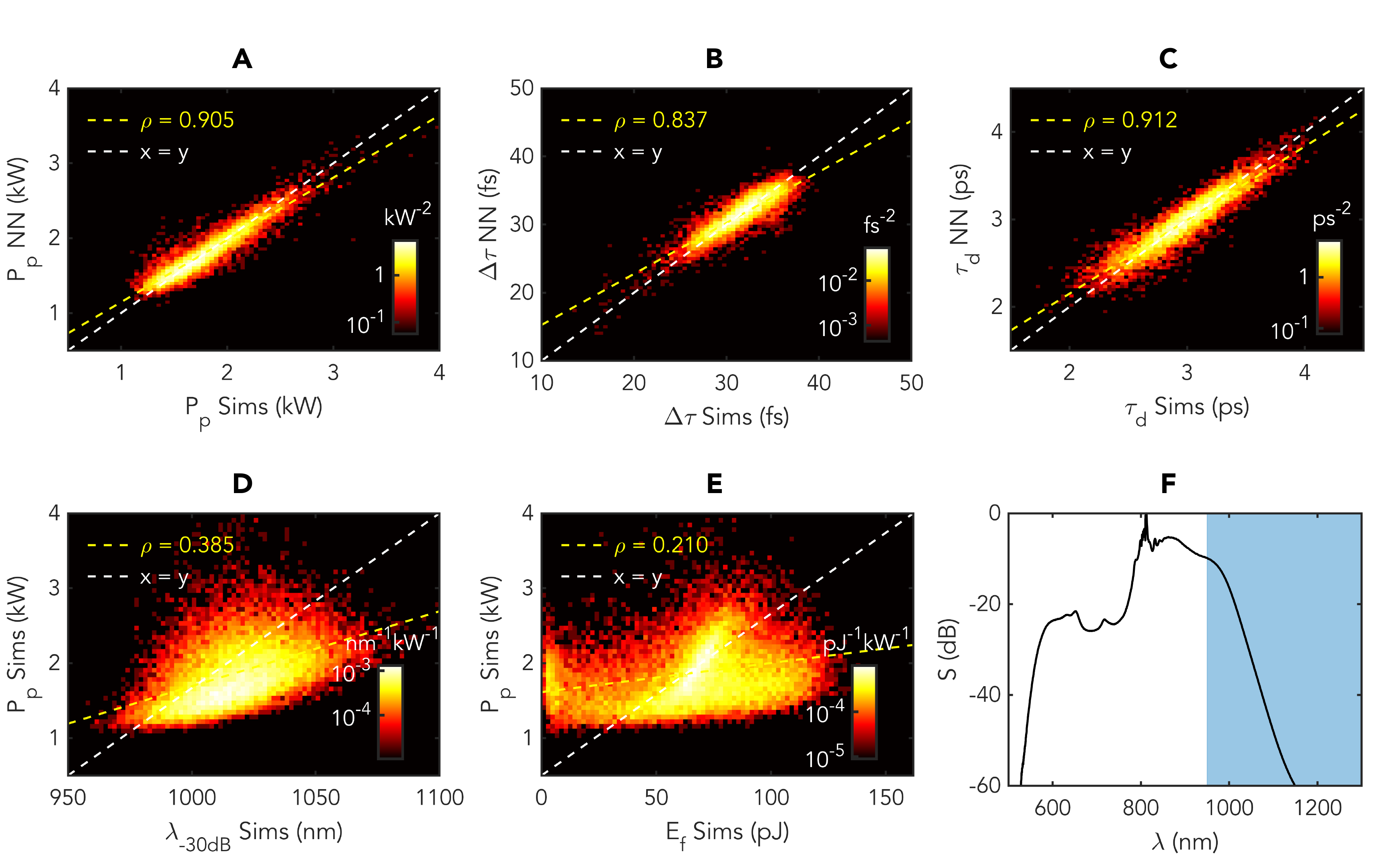}
  \end{center}
  \caption{Results showing prediction of the rogue soliton characteristics by the neural network for an ensemble of 10,000 SC ensemble generated with identical input pulses (except for the noise seed). (A) compares the predicted maximum intensity of the most red-shifted soliton with the exact value simulated time-domain profiles. (B) compares the predicted duration $\Delta\tau$ of the solitons with the exact value from the simulations. (C) compares the predicted delay $\tau_d$ of the solitons with the exact value from the simulations. (D) plots the relation between the SC spectral bandwidth and peak power of the most red-shifted soliton. (E) plots the relation between the SC energy beyond 950 nm corresponding to the shaded area in (F)) and peak power of the most red-shifted soliton. (F) shows the average spectrum of the data ensemble. For all sub-figures, the logarithmic histograms show the normalized density of points grouped into bins of constant area. The dashed white line in each case marks the 1-to-1 correspondence and the yellow dashed line is a linear fit to the predictions along with the Pearson correlation coefficient $\rho$.}
 \label{fig:novar}
\end{figure}

After the training, the NN is tested using an independent ensemble of 10,000 simulated single-shot SC spectra that were not used in the training phase. The results are shown in Fig.~\ref{fig:novar}, where we compare the peak power (Fig.~\ref{fig:novar}A), duration (Fig.~\ref{fig:novar}B) and temporal delay (Fig.~\ref{fig:novar}C) of the most red-shifted soliton predicted by the NN from the single-shot SC spectra and the expected (``ground truth'')  value extracted directly from the simulated time-domain profiles. For convenient visualization, the comparison is plotted as a as a false colour representation of a  histogram (using a logarithmic scale) where the predictions are grouped into bins of constant area, and the histogram shows the normalized density of data points grouped into each bin. For all different characteristics, we can see near-perfect clustering around the ideal ``$x=y$'' relation (indicated by the white dashed line) with a Pearson correlation coefficient of $\rho = 0.91$,  $\rho = 0.84$ and $\rho = 0.91$ for the peak power, duration and temporal delay, respectively (yellow dashed line). 

In contrast to the strong correlations obtained using the NN, we also plot the peak power of the most red-shifted solitons against simpler possible ``predictive'' metrics such as the wavelength corresponding to the -30~dB SC bandwidth on the long wavelength edge (Fig.~\ref{fig:novar}D), and the integrated energy in the long-wavelength edge beyond 950 nm (Fig.~\ref{fig:novar}E). We see clearly that the correlation between these simple metrics and the most red-shifted soliton peak power is extremely poor with  correlation coefficients of $\rho = 0.39$ and $\rho = 0.21$, respectively. These results clearly illustrate the trained NN's capability to accurately predict the characteristics of the most red-shifted soliton from single shot SC spectra without any spectral phase information, and its superiority compared to simpler metrics such as the spectral bandwidth or filtered spectral energy. 

\begin{figure}[t]
  \begin{center}
  \includegraphics[width=\textwidth]{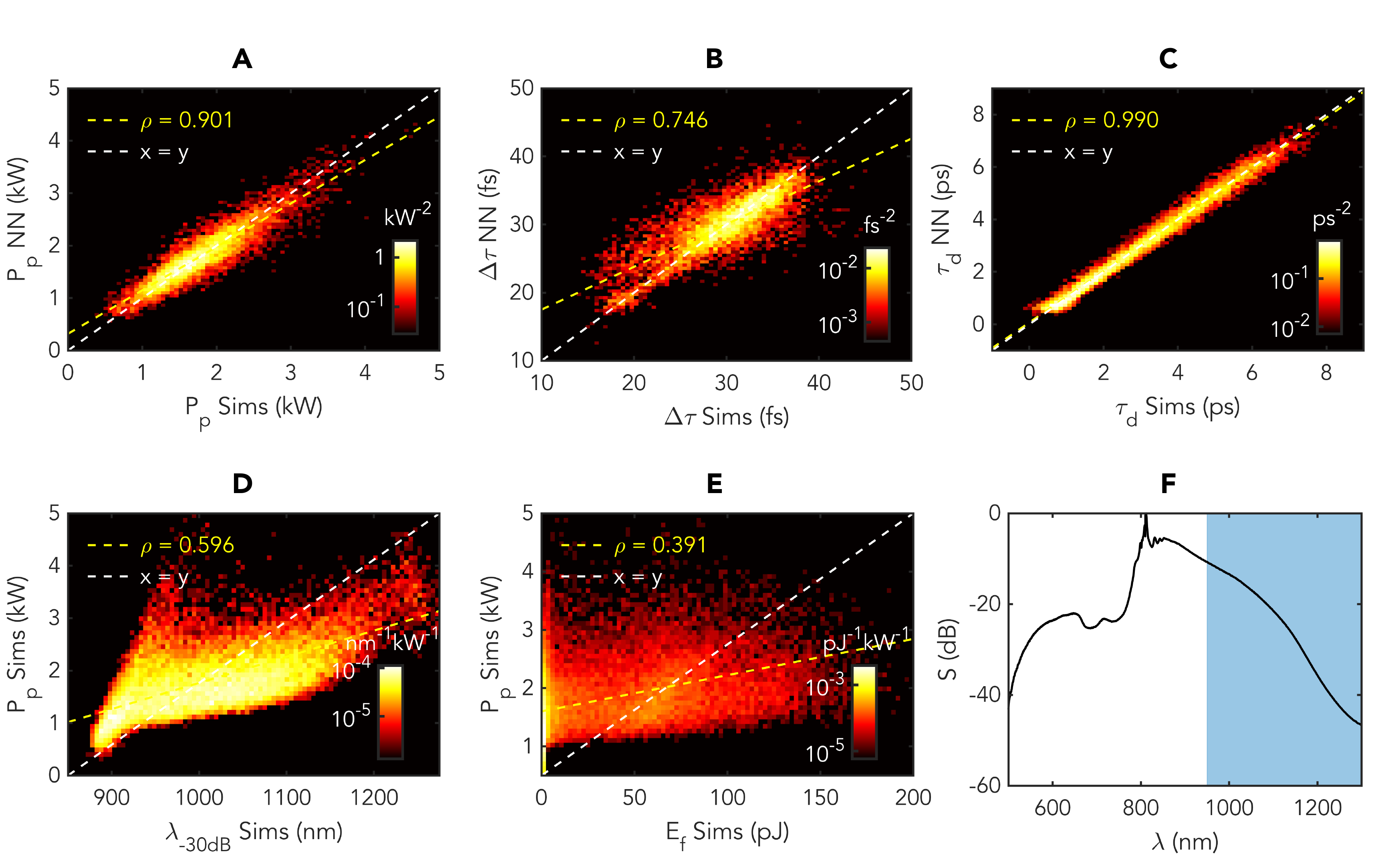}
  \end{center}
  \caption{Results showing prediction of the rogue soliton characteristics by the neural network for an ensemble of 10,000 SC ensemble generated with large ($\pm$~50\%) variations in the input pulse peak power and pulse duration. (A) compares the predicted maximum intensity of the most red-shifted soliton with the exact value simulated time-domain profiles. (B) compares the predicted duration $\Delta\tau$ of the solitons with the exact value from the simulations. (C) compares the predicted delay $\tau_d$ of the solitons with the exact value from the simulations. (D) plots the relation between the SC spectral bandwidth and peak power of the most red-shifted soliton. (E) plots the relation between the SC energy beyond 950 nm corresponding to the shaded area in (F)) and peak power of the most red-shifted soliton. (F) shows the average spectrum of the data ensemble. For all sub-figures, the logarithmic histograms show the normalized density of points grouped into bins of constant area. The dashed white line in each case marks the 1-to-1 correspondence and the yellow dashed line is a linear fit to the predictions along with the Pearson correlation coefficient $\rho$.}
  \label{fig:var}
\end{figure}

To further evaluate the performance of the NN, we ran a series of additional tests where the NN was trained from an ensemble of SC spectra with large variations in the input pulse parameters. More specifically, we added a uniform $\pm$~50\% variation in the input pulse peak power and duration from 200~W to 600~W and from 1~ps to 3~ps such that the training ensemble now contained a large variety of SC development scenarios from essentially pure modulation instability to broad octave-spanning spectra. Using an ensemble of 20,000 simulations, the neural network was again trained to relate the temporal properties of the most red-shifted soliton with the full SC spectral intensity profile, and the network was tested on a separate set of 10,000 simulations not used in the training phase. The comparison between the peak power, duration and delay of the most red-shifted soliton predicted by the NN from the single-shot SC spectra and the ground truth value extracted directly from the time-domain profiles are shown in Fig.~\ref{fig:var}. Despite the large variation in the propagation dynamics, the peak power and delay are again predicted with near-perfect accuracy with a correlation coefficient of $\rho = 0.90$, $\rho = 0.75$ and $\rho = 0.99$, respectively. For completeness, we also plot the most red-shifted soliton peak power vs. the SC spectral bandwidth and energy beyond a wavelength of 950 nm. Again, one can see how the NN performs extremely well, and again with a prediction capacity far superior to the use of a simpler metrics ($\rho = 0.60$ and $\rho = 0.39$, respectively).

\section*{Conclusion}
We have applied machine learning to the analysis of supercontinuum generation in the long pulse regime. Using a feed-forward neural network trained on numerical simulations of the GNLSE, we have shown how the temporal characteristics of solitons with extreme red-shift (peak power, duration and temporal delay) can be predicted with high accuracy based only on single-shot SC spectral intensity profiles without any spectral phase information. The network is able to accommodate and maintain high accuracy for a wide range of regimes, with far superior performance when compared to using spectral metrics such as bandwidth or energy over a specific wavelength range. Our results expand previous studies of instabilities using machine learning, with potential to overcome the limitations of current experimental techniques. 


\section*{Methods}

\subsection*{Numerical modelling}
Numerical modelling is based on the generalized NLSE that describes the propagation of the envelope of an optical field \cite{dudley2006}. We model the propagation of 2~ps duration (FWHM) and 400~W peak power hyperbolic-secant optical pulse centered at 810~nm. The pulse is injected into the anomalous dispersion regime of a 85~cm-long photonic crystal fiber (similar to NKT Photonics NL-PM-750) with zero-dispersion wavelength (ZDW) at 750~nm. The fiber has a second ZDW beyond 1200~nm. The nonlinear coefficient of the fiber is $\gamma$~=~0.1~W\up{-1}m\up{-1} and the Taylor-series expansion coefficients of the dispersion at 810~nm are 
$\beta_2$~=~-1.24$\times$10\up{-26}~s\up{2}m\up{-1},
$\beta_3$~=~8.94$\times$10\up{-41}~s\up{3}m\up{-1},
$\beta_4$~=~-2.54$\times$10\up{-56}~s\up{4}m\up{-1},
$\beta_5$~=~-7.01$\times$10\up{-70}~s\up{5}m\up{-1},
$\beta_6$~=~2.28$\times$10\up{-84}~s\up{6}m\up{-1} and
$\beta_7$~=~-2.21$\times$10\up{-99}~s\up{7}m\up{-1}.
The simulations use 16384 spectral/temporal grid points with a temporal window of 20~ps. Noise is added in the frequency domain in the form of a one-photon-per-mode with random phase. An ensemble of 30,000 simulations corresponding to different input noise seeds was generated. The ensemble was split between two sub-ensemble of 20,000 and 10,000 realizations used for the training and testing, respectively. For the generalization of the training, both the pulse duration and peak power were randomly and uniformly distributed with $\pm$50\% variations from the nominal values above, resulting in dynamics from pure MI to octave-spanning SC. 

\subsection*{Deep learning}

The neural networks relies on supervised learning where one has the knowledge of the relation between the input ${\bf X}_n$ and output $Y_n$ of a specific system \cite{schmidhuber2015}. The training is then based on feeding a large number of distinct examples input and output pairs to create a predictive model that minimizes the prediction error $\varepsilon$ between the desired $Y$ and predicted $Y^*$ values. Here we a training set of 20,000 simulations, where the input is a spectral intensity profile ${\bf X}_n$ (n=1...20,000) and the output $Y_n$ is a temporal characteristics associated with this spectral intensity profile (most red-shifted soliton peak power, duration or temporal delay). The input spectra are pre-processed to a resolution of 1~nm to reduce the computational load in the training process and allow the possibility for reasonable future experimental applications such as in Ref.\cite{narhi2018}.

Thus, the input consists of 801 uniformly distributed wavelength bins from 500~nm to 1300~nm. The input is sequentially fed through the NN with each of the layers operating on the data to yield the desired output at output layer. The connections between the nodes on following layers are weighted and the output of each node is computed as a weighted sum from the output of the previous layer.
Additionally, an adjustable bias term is included for the sum and followed by a nonlinear activation function to yield the output of the node.

The NN consists of the input layer, two fully-connected hidden layers with 80 and 20 nodes, respectively, and a single output node.
The output of a generic $k$th layer $\mathbf{h}^{(k)} \in \mathbb{R}^M$ can be calculated by
\begin{equation}
    \label{eq:weigthed}
    \mathbf{h}^{(k)} = f(\mathbf{g}^{(k)}) = f(\mathbf{W}^{(k)} \mathbf{h}^{(k-1)} + \mathbf{b}^{(k)}),
\end{equation}
where $\mathbf{W}^{(k)} \in \mathbb{R}^{M \times D}$ is a matrix of weights between the layers $k-1$ ($D$ nodes) and $k$ ($M$ nodes).
The vector $\mathbf{b}^{(k)} \in \mathbb{R}^{M}$ contains the bias terms for each node in layer $k$ and $f()$ is the activation function.
In this work, hidden layers were connected by hyperbolic tangent sigmoid activation function \mbox{$f(x) = 2/[1 + \exp (-2x)] - 1$},
and a single node with linear activation function was used for the output layer.
The output for a single generic node in layer $k$ can be written as
\begin{equation}
    h_i^{(k)} = f\left( \sum_{j=1}^{N_{k-1}} w_{ij}^{(k)} h_{j}^{(k-1)} + b_{i}^{(k)}\right).
    \label{eq:singleou}
\end{equation}
Here $w_{ij}^{(k)}$ are the weights between nodes $i$ and $j$ in layers $k$ and $k-1$, respectively.
Variable $b_{i}^{(k)}$ is the bias term associates with node $n_{i}^{(k)}$.
The summation includes all the $N_{k-1}$ nodes in layer $k-1$.
We use mean squared error function
\begin{equation}
    \varepsilon = \frac{1}{N} \sum_{n=1}^N (Y_n - Y_n^*)^2,
    \label{eq:error}
\end{equation}
where $N$ is the number of samples, and $Y_n$ and $Y_n^*$ are the target and predicted outputs, respectively.
For the training, a conjugate gradient back-propagation \cite{fletcher1964} is used.
The weights and biases are adjusted relative to their current state according to their partial derivatives respect to the error function.
The "speed" of the adjustment determined by the learning rate or step size $\eta$.
The adjustment to weight $w_{ij}^{(k)}$ is taken towards the negative gradient of the error function, given by \mbox{$\Delta w_{ij}^{(k)} = -\eta \partial \varepsilon / \partial w_{ij}^{(k)}$}.
Once all of the input and output pairs (${\bf X}_n, Y_n$) in the training set have been passed through the network one \emph{epoch} has passed.
The networks were trained for 500 epochs until convergence.
After training, the NN is tested with a separate set not used in the training phase to evaluate the performance of the network.

\bibliographystyle{ieeetr}
\bibliography{RSml}



\section*{Acknowledgements}
GG acknowledges the Academy of Finland (298463, 318082, Flagship PREIN 320165). LS acknowledges the Faculty of Engineering and Natural Sciences graduate school of Tampere University. JD acknowledges the French Agence Nationale de la Recherche (ANR-15-IDEX-0003, ANR-17-EURE-0002). 

\section*{Author contributions statement}
LS performed the numerical simulations. GG and JD supervised the project. All authors contributed to the data analysis and writing of the manuscript. 

\section*{Additional information}
\textbf{Competing interests:} The authors declare no competing interests.

\end{document}